# 基于空间句法和神经网络方法的旅游景点道路拥挤度评估模型研究----厦门鼓浪屿为例

**苏清木　　肖晶晶　　余宣和　　陈翔**

# Research on Evaluation Model of Road Congestion of Tourist Attraction Based on Spatial Syntax and Neural Network Method--A Case of Gulangyu Island, Xiamen，China

**【摘要】**快速发展的城市化和历史古镇的旅客大量的增加，使得历史古镇的保护经常被忽略，也对旅游空间的可持续重组提出了挑战。在现今的旅游开发中，土地使用与交通规划却是趋于分离的两项任务，这造成了道路不合理分配的问题。我们知道，道路是景点通行的载体，道路的拥挤情况会对周边景点造成破坏。但现今的研究中，仍很少将街道网络结构、景点分布和行人运动之间进行综合研究。为了较准确的预测行人的流量，了解旅游空间与行人之间的互动关系，本文利用空间句法和神经网络方法建构了旅游道路拥挤度的评估模型。该模型充分发挥神经网络方法和空间句法的优势，比如，神经网络方法能够对景点权重进行客观动态的赋值，并且能够通过训练推估出其他景点的权重；空间句法可以对中国鼓浪屿的街道可达性进行分析，可以很清晰地了解道路之间的连接关系；接着我们利用数学公式将道路网络结构和景观吸引点进行有效结合，这可以在街道网络结构、景点分布和行人运动之间的对应性较低且不一致的情况下估算道路拥挤度的能力。我们应用厦门鼓浪屿对其进行了实验，其结果，我们发现，（Ⅰ）鼓浪屿的景点吸引力较高的地区主要分布在岛屿的边沿，有卖门票的几个景点吸引力都达到 0.9 以上；（Ⅱ）空间句法的拓朴模式能够更好预测鼓浪屿游客的步行结果；（Ⅲ）鼓浪屿的道路可达性和景点的分布在空间上并没有很大的相关性，但模型能够预测道路的拥挤度，使其更接近事实。我们的研究结果，可以作为未来旅游空间管理的依据，也可以丰富旅游空间的研究。

# 1.前言

历史古镇作为城市历史和记忆的一部分，是城市风格和城市景观基本元素的绝佳代表(Wang, 2012)。但在快速发展的城市化进程中，历史古镇的保护经常被忽略，甚至有些历史街区在旧城区的更新下被完全消失了。而有些古镇为了更多的旅游经济收入和过度追求城市建设的速度与现代化，使得原有的空间形态结构发生了巨变，原有空间形态中特有的场所精神和文化特性逐渐消失(Oppong et al., 2018, Li et al., 2019)。特别是在中国这样一个快速发展的国家，更加追求城市的扩张和经济的发展，这使得历史古镇的保护压力更大(Wu et al., 2018)。随着人们社会文化生活的变化，历史古镇的旅客大量的增加，也使古镇保护压力倍增，因此，保护历史遗产和旅游空间的可持续重组已成为城市规划者的共同议程(Li et al., 2016, McKercher et al., 2005, Wang and Bramwell, 2012, Dela Santa and Tiatco, 2019)。

为了支撑历史古镇的永续发展，古镇的开发必定要对交通路网和城市用地进行有效的整合。但，实际上，土地使用与交通规划却是趋于分离的两项任务，这使得道路路网的可达性和机动性并不能很好的跟土地使用的行为进行有效的连结(Kaiser et al., 2006, Zeng et al., 2019)。就历史景点来说，往往是基于当前的市场对道路网进行评估，而未考虑后期旅游景点的开发，这造成了道路不合理分配的问题。因为旅游景点的开发，会产生的大量的旅客行为，会加重原有道路的拥挤度。就规划者而言，规划方案不仅是为了帮助了解既存环境里的使用问题，同时也是为了下一阶段的设计和引导，并提供问题的答案和修正的方式，以避免类似的错误发生(Hillier, 2007, Li et al., 2017)。在旅游空间规划中，如果未能对旅客行为进行有效的模拟和评估，那么可能就会造成旅游景点空间不合理安排的情况和加重旅游空间拥挤性。我们知道，道路是景点通行的载体，道路的拥挤情况会对旅客的心情产生影响，也会间接影响景点的经济效率，同时过度的拥挤也会对周边景点造成破坏(Saenz-de-Miera and Rosselló, 2012)。本文研究道路网和景点之间的关系，可以作为历史古镇未来调整的依据。

本文中的中国鼓浪屿是现代化城市发展中的一个实例，其具有悠久的历史和人文底蕴，已成为一个越来越受欢迎的国内外旅游目的地。大量的旅客到访（比如，2019 年第一季度就有超过 233.9 万人登岛参观(Gulangyu, 2019)），使得其社区文化、古城保护和岛屿的吸引力日益受到威胁(Lin, 2010, Yu and Liu, 2012)，并且鼓浪屿管理委员会也没有相应的行人拥挤度预测来对旅客进行疏导。因此，需要评估规划道路可能的拥挤度，以便于对旅客进行疏导。在建立道路评估模型方面，以往主要是利用空间句法的方法概念和框架进行建构。大多数研究是以街道网络的中心属性或土地使用作为分析对象，但未能将其进行有效结合考虑。也就是说，他们无法评估一个地点的土地使用情况如何透过街道网络结构影响其他地点的移动(Hillier et al., 1993, Omer and Kaplan, 2017, Raford, 2010)。因此，旅游空间拥挤度的评估需要建构一个复合型的评估模型，对规旅游空间规划拥挤度进行定量的评估模拟，并同时考虑了街道网络结构和景点吸引力对行人的影响。

建构一个复合型的评估的模型。首先，街道网络结构的考虑上，采用的是空间句法。该模型可以很清晰地分析道路之间的连接关系，但该模型受集中度测度类型和分析半径选择之间的影响，因此，文章将对其进行研究，以选择最符合本研究的集中度测度类型和分析半径(Jiang, 2009, Omer and Jiang, 2015, Hillier, 2012)。其次，在景观吸引力考虑上，由于现阶段没有统一的标准，并且每个地区的景点影响因子不同，如果我们根据以往的权重进行评量，就有可能造成对现状理解的偏差，同时,如果采用专家评分的方式（PHA），则可能导致结果过于主观，因此有必要利用其他方法，使结果更加准确。很多研究已表明，神经网络分析方法可以和好的解决上述的问题(Xiao et al., 2017, Zhang et al., 2019, Slimani et al., 2019)。神经网络分析方法可以通过模拟人脑的思考方式对旅游资源建立信息化模型，并利用信息化模型的评估结果，找到这些因素之间的直接关系。它可用来评估和预测实际的旅游景点状况，从而确立景观的吸引力。因此，本文利用神经网络分析，可用来弥补空间句法在分析景观吸引力的权重上的不足。

本文发挥空间句法和神经网络方法的优势，通过空间句法对中国鼓浪屿的街道可达性进行分析，通过神经网络方法对景点吸引力权重进行赋值，并利用数学公式将道路网络结构和景观吸引点进行有效结合。其结果采用 arcGIS 进行空间可视化，从而更好的呈现道路景点的拥挤度。本研究有利于旅游景点未来的引导和方向上的调整。在这过程中本文试图回答下面几个问题：（Ⅰ）历史古城的景点吸引力是什么？景点之间有什么差别？（Ⅱ）道路的可达性如何影响旅客的行为？（Ⅲ）景点吸引力

和道路可达性如何相互影响？在双重影响下，道路拥挤度又是如何变化的？

# 2.研究方法与框架

## 2.1 研究区域

本文选定鼓浪屿作为研究区域。鼓浪屿上有许多历史景点遗迹且是非常著名的旅游岛屿之一。（待补充）

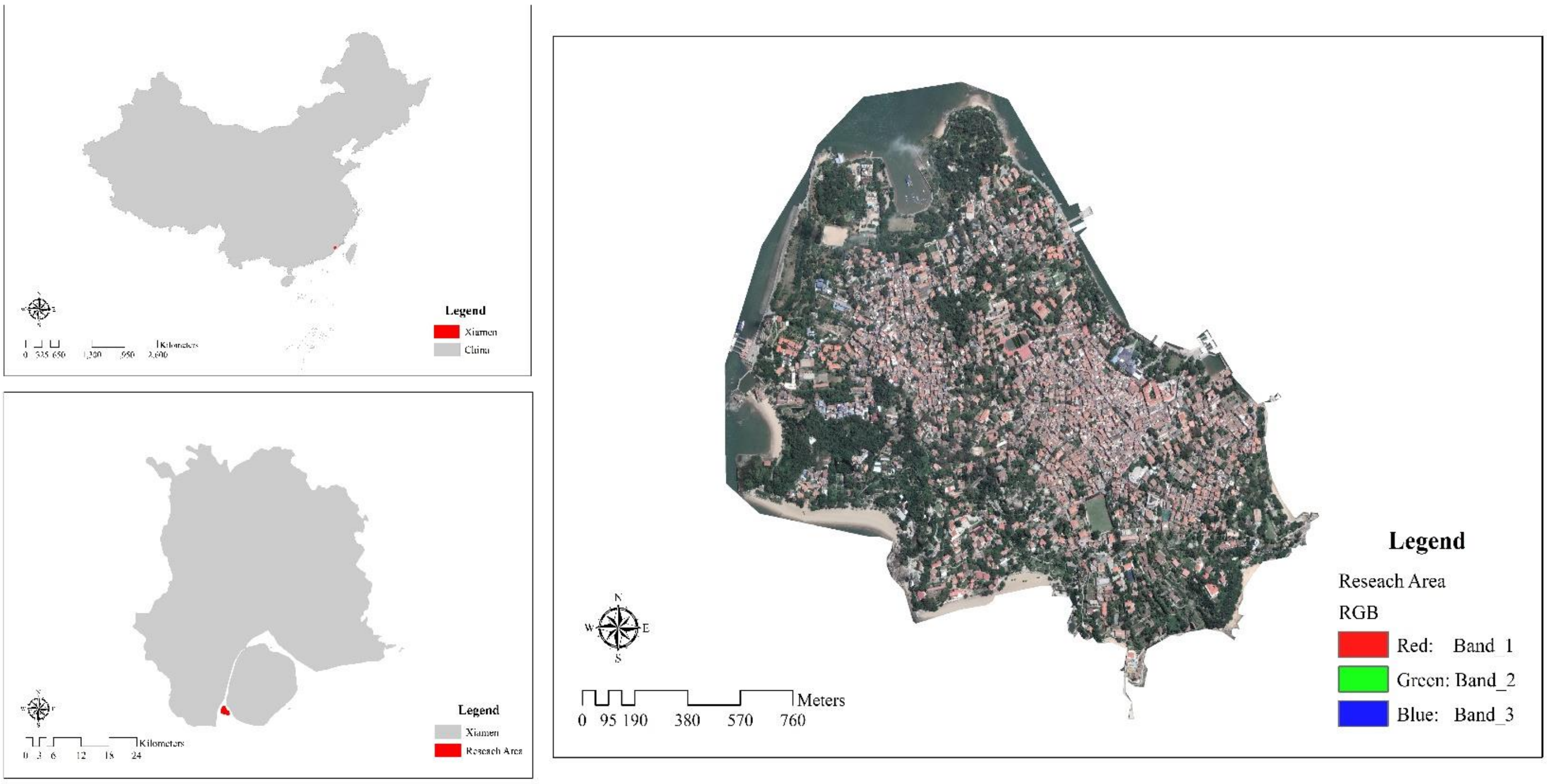

## 2.2 方法

### 2.2.1 神经网络分析

神经网络（ANN）是一种模拟人脑中数据处理方式的数据处理系统(Acheampong and Boateng, 2019)。由于人工神经网络的拓扑结构是基于人类大脑的生物神经网络的结构，因此，其具有并行处理信息技术的能力和根据过去已有的实例样本进行“自学习”和模式识别的能力。此外，人工神经网络可以直接处理不可确定的非线性的参数，这也是它跟统计工具最大的区别(DeTienne et al., 2003)。本研究为了预测景观吸引点的权重，选择具有反向传播（BP）的多层感知器（MLP）以实现 ANN 模型的最佳性能，BP 神经网络是目前应用最为广泛的模型(Karayiannis and Venetsanopoulos, 2013, Sohn, 2014, Krishnaveni and Pethalakshmi, 2017, Li and Fernando, 2016)。本文采用 BP 神经网络分析在于对景点权重进行客观动态的赋值，以便能够动态的观察不同影响要素下，景点的吸引力变化，能够更加智能的根据现状变化进行评估。

BP 神经网络是一种单向传播的多层前向神经网络,分为输入层、隐含层和输出层,层与层之间采用全连接方式,同一层单元之间不存在相互连接，并且每个链接都与权重相关联，能够通过交替的权重值进行学习（图 2）。

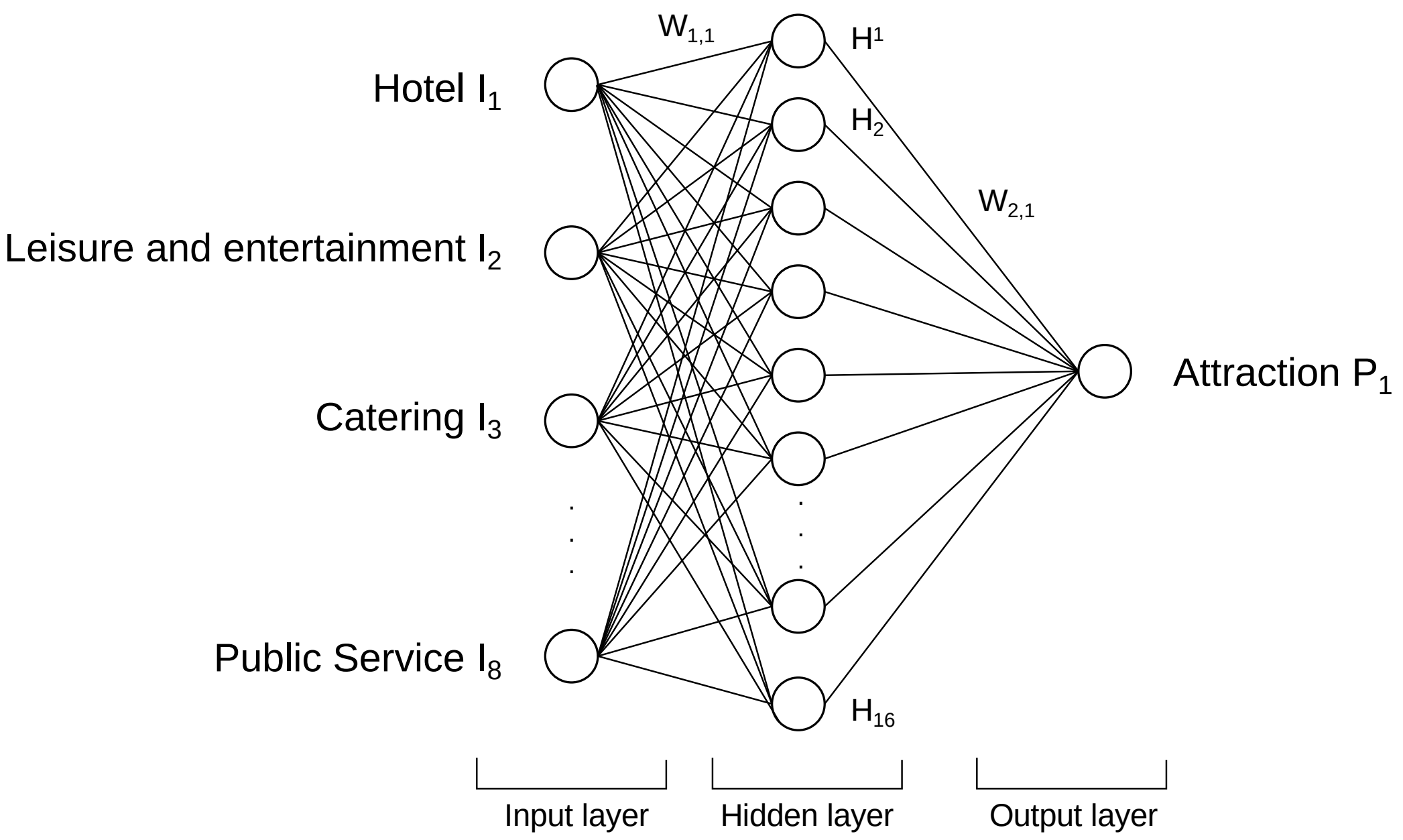

图 2 神经网络输入层、隐含层和输出层之间的关系

输入层由多个输入节点组成用于接收用于预测、分类和回归的输入数据。由于在训练 ANN 时高维数据的严格计算，需要对数据进行标准化和归一化(Boateng et al., 2019)。因此本研究，为了消除各景点影响因子的数值标准不一对输出结果造成不良的影响，采取归一化的处理方式，即将数据转变成[0,1]之间的数据，其公式如下：

$$F_i = \frac{(x_i - x_{i\,\min})}{(x_{i\,\max} - x_{i\,\min})} \quad (1)$$

其中，$F_i$ 为 $x_i$ 的标准值，$x_i$ 指的是输入数据，$x_{i\,\min}$ 指的是输入数据里面最小的值，$x_{i\,\max}$ 则指输入数据里面最大的值。

隐藏层是用来处理输入层非线性数据的内部替代参数，由非线性激活函数组成，例如 sigmoid 函数，hardlimit 函数，径向基函数和三角函数。目前 sigmoid 函数是 BP 神经网络最常用的激活函数(Acheampong and Boateng, 2019, Alvanitopoulos et al., 2010)，Sigmoid 函数的公式如下：

$$Y_{ij} = \frac{1}{1+\exp(-Y_j)} (j = 1,2,3,4,5) \quad (2)$$

其中 $Y_{ij}$ 是 sigmoid 函数，$Y_j$，隐含层结点，是 $\exp(-Y_j)$ 是指数函数。

输出层为所需输出数据格式的大小，是用来回答所研究的问题。在理论上，输入层有 n 个神经元，则隐含层有 2n 个神经元,输出层有 m 个神经元，可以实现网络拓扑结构的任意连续映射，因此采取的结构为 $n \times 2n \times m$ 的神经网络模型(Karayiannis and Venetsanopoulos, 2013)。本研究共选择了 8 个具体标准，因此，在建立景点吸引力模型的时候采取结构为 $8 \times 16 \times 1$ 的神经网络模型,即 8 个具体影响因子作为输入矢量,景点地得分为唯一的输出矢量,隐含层节点数量为 16。

**景点影响因子的考虑**

在计算神经网络时，需要确定景点的影响因子。景点是吸引游客的关键因素，也是当地社区的资源，作为旅游产品和体验不可或缺的要素，游客景点是旅游活动的重点，是游客前往目的地停留和停留的原因(Leask, 2010)，其与周边景点的联动，可以整体提升景点的知名度。同时，住宿的社会和经济吸引力极大地影响了旅游目的地选择的扩展，使得旅行频率、停留时间和活动范围有所增加(Tussyadiah and Pesonen, 2016)。当旅行时所有旅客都需要吃饭，他们的需求根据对食物的态度和

承诺而有所不同，而每位旅客与食物相关的行为一般都指向餐馆，有些人认为食物对目的地选择至关重要(Björk and Kauppinen-Räisänen, 2016)。另外，一个地区基础设施建设的好坏，会促进或制约了该地区旅游业的发展，研究表明景观和基础设施也与访客选择旅游地有着直接或间接相关(Grünewald et al., 2016)。总所周知，一个地区旅游业的发展相辅相成的会带动经济的发展，比如会带动周边的酒店住宿、餐饮、交通、商铺还有其他基础设施等的发展，而反之，景点周边的酒店、餐饮、商铺、交通以及其他基础设施情况也会在一定条件下对游客选择景点目的地产生一些影响。综上本文根据这些相关研究和鼓浪屿岛的内部条件，设计了会影响景点吸引力的 8 个影响因子，包括周边景点数、住宿数量、餐饮数量、以及把基础设施具体分为商家数量、休闲娱乐点数量、公共服务点数量、生活服务点数量、交通设施点数量。

**计算景点缓冲区**

确定了景点的影响因子后，需要对其进行赋值。在具体标准的赋值上，本文以厦门鼓浪屿作为研究区域，选取了研究区域内 229 个景点作为观光点，并以这 229 个景点为中心，200 米为半径计算缓冲区。其中 200 米的半径是综合区域大小和景点大小决定的。缓冲区的定义如下：

$$p=\{x \mid d(x,k_t)\le r\} \quad (3)$$

其中 $k_t$ 表示观光点的坐标，$x$ 表示缓冲区内的 POI 地点，函数 $d$ 表示空间上的距离，所以 $p$ 点是观光点 200 米半径内 POI 的集合。各个景点的影响因子的具体数量见附表。

**BP 神经网络的训练**

对景点影响因子的赋值后。本研究将景点吸引力指标作为输入矢量，将到访鼓浪屿的旅客作为目标输出，通过隐藏层和输出层的计算，不断获取各层单元的参考误差，一直训练到网络输出的误差在可接受的程度内或达到设定的学习次数为止(Hillier, 2007, Xiao et al., 2017)，其具体步骤如图 3。具体来说，我们以景点影响因子作为自变量，以景点的门票数作为因变量。由于鼓浪屿只有对日光岩、菽庄花园、鼓浪屿风琴博物馆、皓月园、鼓浪屿国际刻字艺术馆进行门票记录，因此，我们选定这几个景点 2019 年上半年每一个月的门票数作为我们神经网络的训练样本，进而可以推估出所有景点的权重。这样做的好处在于，对于其他不需要门票的景点，也能得到景点的权重。

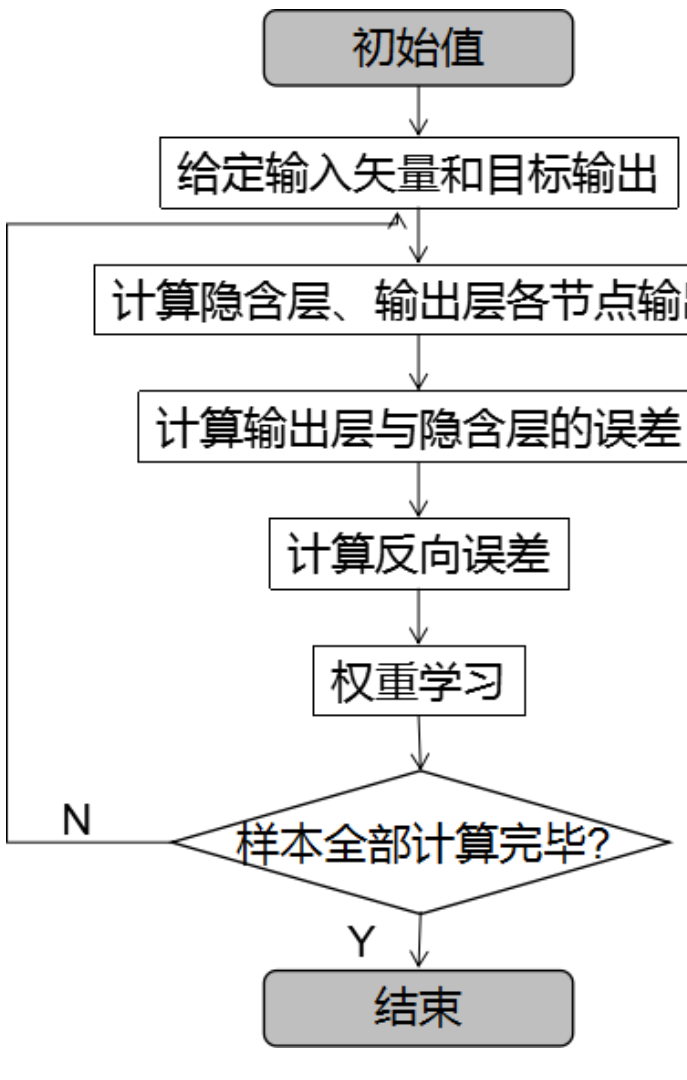

图 3 神经网络具体步骤

## 2.2.2 空间句法分析

空间句法是主要用来研究空间组织和人类社会之间关系的一种理论和方法。目前空间句法在建筑设计和城市规划领域得到了广泛的应用，主要因为空间句法理论可以预测建筑设计和城市规划成果在未来的效果，从而使得规划管理者在进行开发决策时可作出明智的决定(Karimi, 2018)。

空间句法的分析方法主要有轴线法、凸空间法、可视域法及线段分析法(Klarqvist, 2015)，本文研究的主题是旅游空间规划，其主要采用的是线段分析法。线段分析法可分解为拓扑、角度、距离等三种模式，这些模式的差异在于对“最短路径”的定义不同(Bafna, 2003)，具体定义差别如下：

（1）拓扑模式（Topological）: 线段中的拓扑分析模式等效于轴线分析，即两点之间线段转折次数最少的路径，亦或者是经过其他线段数最少的路径。

（2）角度模式（Angular）:角度模式是线段分析中最常用的分析模式，其定义为两线段间综合

折转角度最小的路径为最短角度路径。

（3）距离模式（Metric）：距离模式是指两线段间距离最短路径。

同时，在过往研究中，空间句法包含多种度量标准，且这些标准目前已广泛引用于测量环境中的配置属性(Chiang and Li, 2019)。而空间句法的行人流量模型通常基于两种类型的集中度测量值，包括整合度和选择度，不同的选择会对结果产生差异(Hillier and Iida, 2005, Omer and Kaplan, 2017, Hillier, 2007)。

（1）整合度通常用于衡量每个片段与所有其他片段之间的可访问性，以及它作为目的地被选择的可能性有多大，举例来说，人通常会把商店放在一个容易接近的所有其他部分的位置，因为人们会通常多半进行短途旅行，较少进行长途旅行。整合度能够体现城市中心的层级差异。

（2）选择度则是在衡量每个分段的直通运动潜力，也可以说是这个片段作为旅行必经之处的能力。选择度可以作为评估路网层级的依据。

**测量半径与分析方式选择**

本研究以空间句法分析街道网络可达性，并使用 Depthmap 软件进行街道网络中心度测量。古浪屿古城范围由长边 2000 米、宽 1000 米区域所构成，大部分街巷以步行为主，并以船运作为联外交通的运输方式，故整个鼓浪屿可视为一个封闭区域进行分析。根据空间句法分析的理论基础，中心度测量将会受到半径选择方式而影响而有所不同，因此，本研究将使用三种距离类型评估，包含距离、角度、拓朴等距离半径。此外，由于半径大小的选择也会影响中心度测量的结果，本研究考虑古城范围长度，以半径 R 为 50,100,200,300,400,500,600,800,1000 及 n，并以米为距离测量单位，作为评估半径选择的基础依据。因此，本研究的半径选择方式是综合考虑不同半径类型及半径大小对中心度测量所造成的结果。通过不断试验以求最优拓扑半径、角半径和距离半径的数值，并促成拥挤度模型最优半径解。

考虑本研究的研究目的是藉由历史古镇的道路可达性与景点影响因子间的关系，来评估道路拥挤度，进而在未来有机会对古城的旅游轴线进行调整，因此，了解景点与景点间必经要道的路网分析是相当重要的，所以比起以整合度为分析手段，选择度的分析是更为合适的。后续本研究将会以距离选择度、角度选择度与拓朴选择度进行街道拥挤度分析，并比较两两之间的 $R^2$ 值，根据分析结果选出适当的分析半径类型及半径大小，并将选择度的数值结果标准化，以此提取各条街道数据，作为后续以 GIS 软件研究景点吸引力与街道网络结构对行人影响时的街道拥挤度参数。

### 2.2.3 拥挤度评估模型建立

本文根据上文神经网络的分析方法和空间句法的分析结果，进行拥挤度模型的构建。为了将其进行有效的连结，首先，我们需要对空间句法的结果和景点吸引力影响因子权重的结果进行标准化处理，并通过 ArcGIS 软件将其变成矢量图层。其次，利用 ArcGIS 软件计算景点到每条道路线段的最短距离，并进行同一景点到不同道路线段的最短距离累加和计算，得到所有景点到各个道路最短距离之和，我们将其定义为 X。最后，通过数学公式将景点吸引力和道路可达性进行连结。具体公式如下（4）和（5）：

$$K = \frac{A}{\ln X} \quad (4)$$

K 为景点对道路拥挤度的影响权重；A 为神经网络得到的景点吸引力影响因子权重；$\ln X$ 表示景点吸引力对道路的影响随着距离的增加呈对数函数锐减，之所以这样考虑是因为，随着景点与道路的距离的拉大，景点吸引力对道路拥挤度的影响趋近于零，并且距离越近，影响越大。在得到点对道路拥挤度的影响权重（K）后，道路的拥挤度模型可表示为：

$$C(L_i) = \sum_{j=1}^{m} K(A_j) \otimes W(L_i),\ j = 1,2 \ldots m, i = 1,2 \ldots n \quad (5)$$

其中 C 为以道路 $L_i$ 的拥挤度值，其值越大，代表道路拥挤度越高；K 为景点 $A_j$ 对道路拥挤度的影

响相对权重参数；m 为 $L_i$ 路段的景点数；W 为 $L_i$ 路段的空间句法拓扑结构中心度测量的相对权重参数。

根据此公式，我们知道，选择某一路段作为目的地的概率与该路段的景点吸引力成正比和该路段的道路可达性成正比。通过其拥挤度模型有利于我们对鼓浪屿岛上的路段进行分析与评估，有利于对鼓浪屿的路网进行空间规划。

## 2.3 研究架构

研究框架设计上，本文基于空间句法和神经网络分析方法的框架下，提出了评估旅游道路拥挤度的评估模型（图 4）。本评估模型首先是对空间规划拥挤度评估模型参数进行设计。因为以往的研究都是将街道网络结构和景点吸引力作为独立实体来处理，难以表现景点的吸引力和街道网络空间的整体观点。因此本文先用神经网络分析方法来订定各景观节点的权重，应用 Matlab 软件，并按照相关理论，采用对数 Sigmoid 的 logsig 函数为输入层与隐含层之间的传递函数，线性 purelin 函数作为网络的隐含层与输出层训练函数选用的函数，来计算影响因子在标准化指标之后经过神经网络分析方法得出的各景观节点的权重。

其次，在街道网络可达性上，采用 depthmap 软件进行空间句法街道可达性分析，其理论基础是中心度的测量，而中心度的测量又受测量半径与分析方式选择的影响，因此会有多种结果产生，再根据研究的假设和研究区域的特征，对其多种结果进行进一步的筛选，使其更符合本研究的假设。

最后，在计算出各景点的吸引力和街道网络空间的可达性后，我们可初步得出景点空间的结构特征模型，接着我们应用 GIS 软件构建了基于景点的吸引力和街道网络空间（拓扑结构）的模型，该模型考虑的是景点对道路的拥挤度的影响是呈对数锐减的，通过数理模型将其集成起来，集成表示给定路段在网络中的可访问性(即它的移动潜力)，从而来估计旅游道路空间的拥挤度。得出景点与街道的拥挤度模型后，我们对模型进行了比较，比较其相对于只考虑景点吸引力或街道网络空间要素之间的不同之处。根据道路拥挤度的分析结果，我们可以判断其线路决策与人流的冲突点、景观吸引力对道路的压力等，来评估街道网络空间的优劣程度，并可根据模型的评估结果来对现状进行调整。

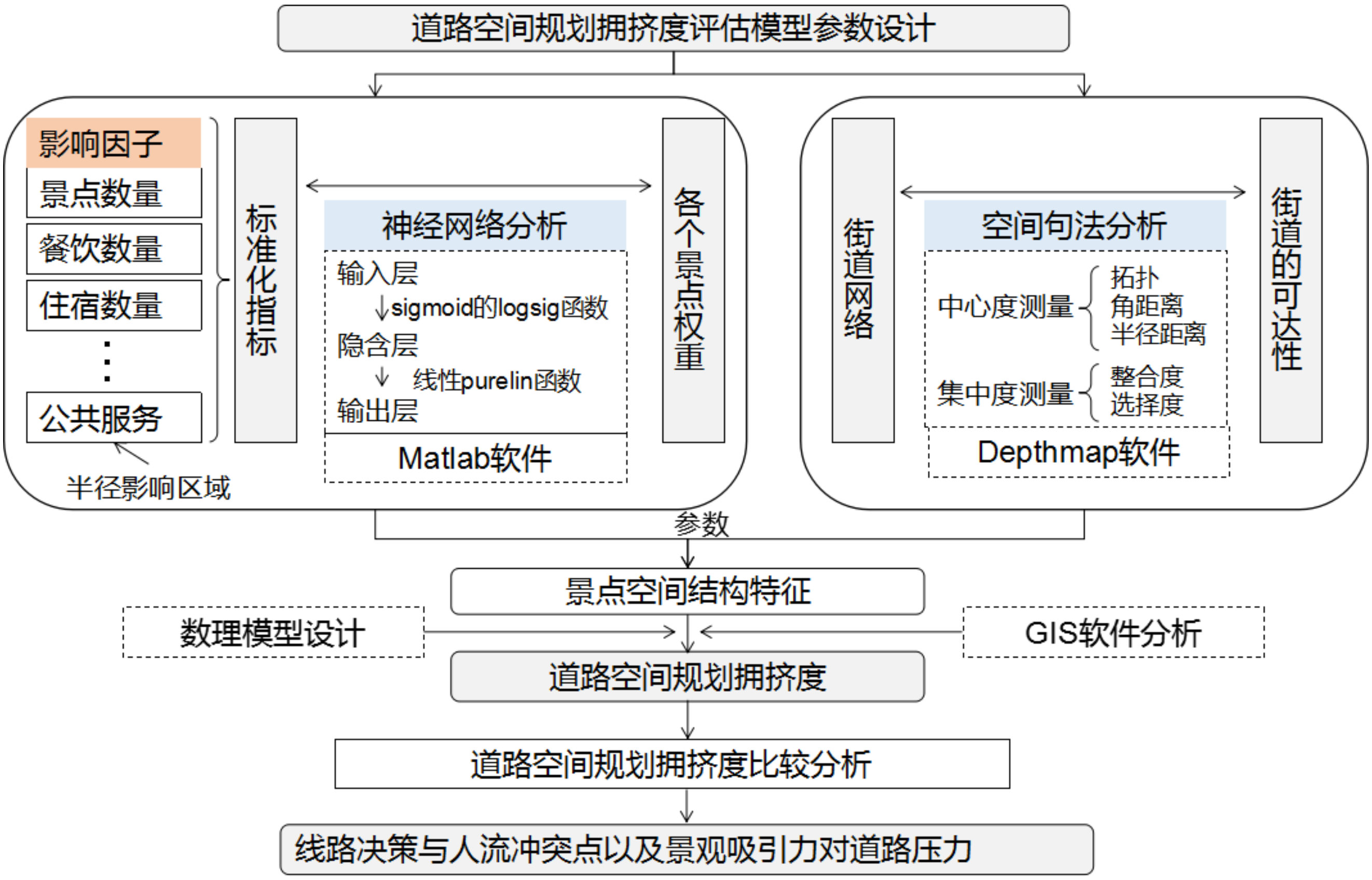

图 4 研究框架图

## 3. 案例研究

### 3.1 景点影响因子的神经网络分析结果

本文的重点是运用神经网络模型进行景点参数设计，将影响景点吸引力的具体标准作为输入节点、将所选景点得分作为网络输出，从而得到各景点的权重。

在神经网络模型的训练与评分上，按照相关理论，采用对数 Sigmoid 的 logsig 函数为输入层与隐含层之间的传递函数，线性 purelin 函数作为网络的隐含层与输出层；训练函数选用的是 trainlm 函数，设定目标误差为 0.00001。首先对建立的网络进行了初始化，然后输入数据经过训练，经过多次迭代后网络误差达到可接受的范围。从图 5 看出，训练的 R 值都超过 0.9，可见，所建立的网络达到稳定。这表明已建成的 BP 神经网络已经可以用于景点吸引力的分析和评估。得出的最后评分结果见附表。

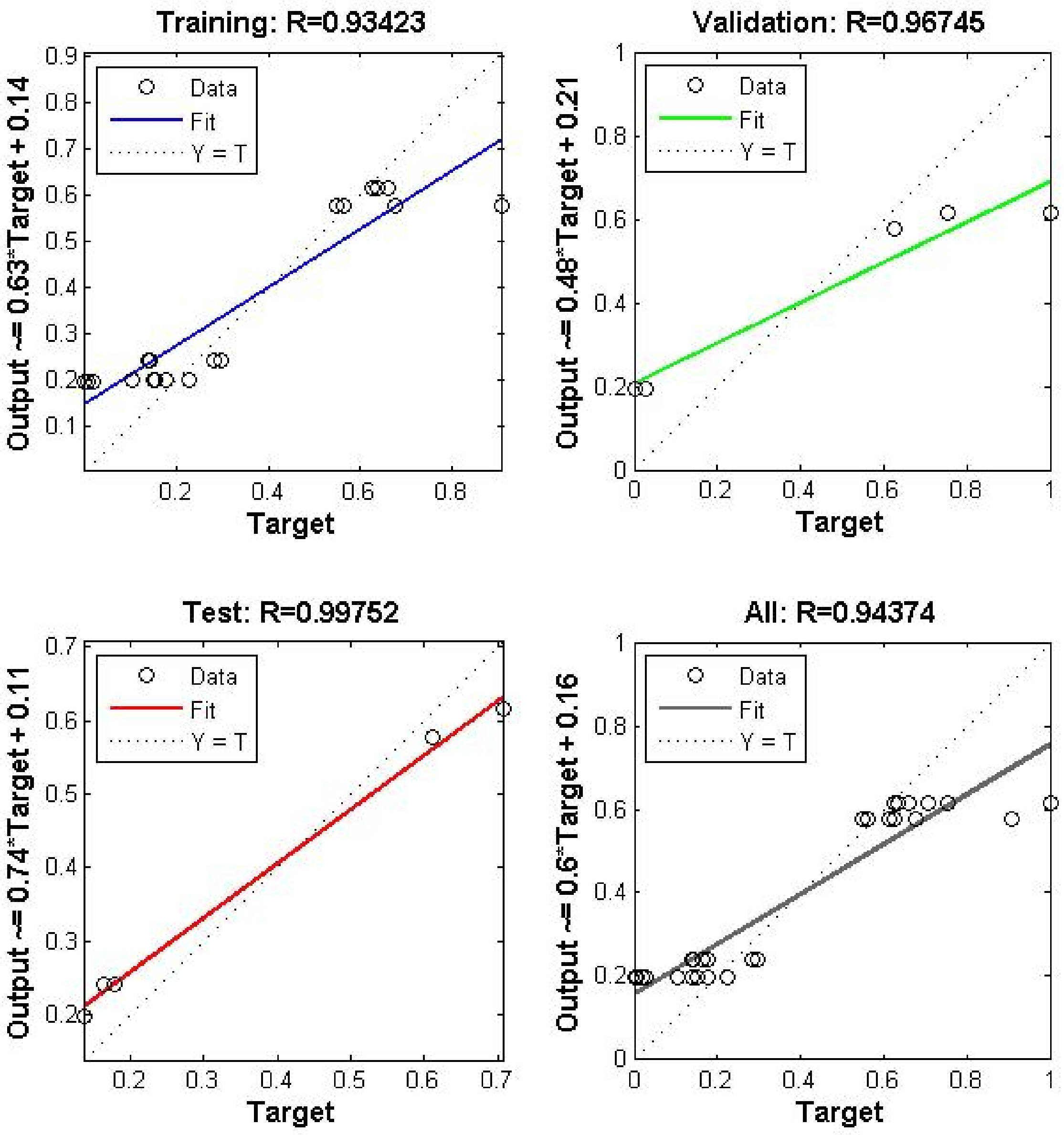

图 5 网络误差的标准值

景点吸引力的分布特征上，本文根据神经网络分析所得到的各景点权重值，将其进行十等分，并重新导入 ArcGIS 软件中按其属性分类，得到景点影响因子权重分类图（图 6），颜色越深越红，代表景点权重值越大。根据景点 200 米缓冲区内的 8 个影响因子的神经网络分析的结果表明，首先，景点存在集结的现象，高权重的景点主要分布在岛屿的边沿，并且主要分布在 A,B,C,D 四个结廓中，这也

间接反映了沿海的旅游资源较丰富，旅游开发程度高，同时也反映旅游景点有聚集开发的效应，这样的目的可能是为了共同开发，减少成本，也可以共同提高旅游景点的吸引力。其次，我们从表 1 进一步发现，有卖门票的 5 个景点（日光岩（0.981）、菽庄花园（0.978）、鼓浪屿风琴博物馆（0.975）、皓月园（0.970）、鼓浪屿国际刻字艺术馆（0.918））的景点权重都处于高水平状态，都达到 0.9 以上，可见这几个景点的旅游资源丰富，是鼓浪屿较有代表性的景点。最后，我们发现城镇中心区的旅游景点的影响权重普遍不高，这也反映出中心区的旅游资源较弱，对旅客较没有吸引力，间接说明了中心区主要为岛上的居民提供服务，而不是以旅客服务为主，不能很大水平的满足游客的观光、消费、娱乐、休闲等多种需求。综上所述，游客往景点权重值大的地方聚集，会造成其附近街道的拥挤度提高，从而加重通往景点道路的负担。

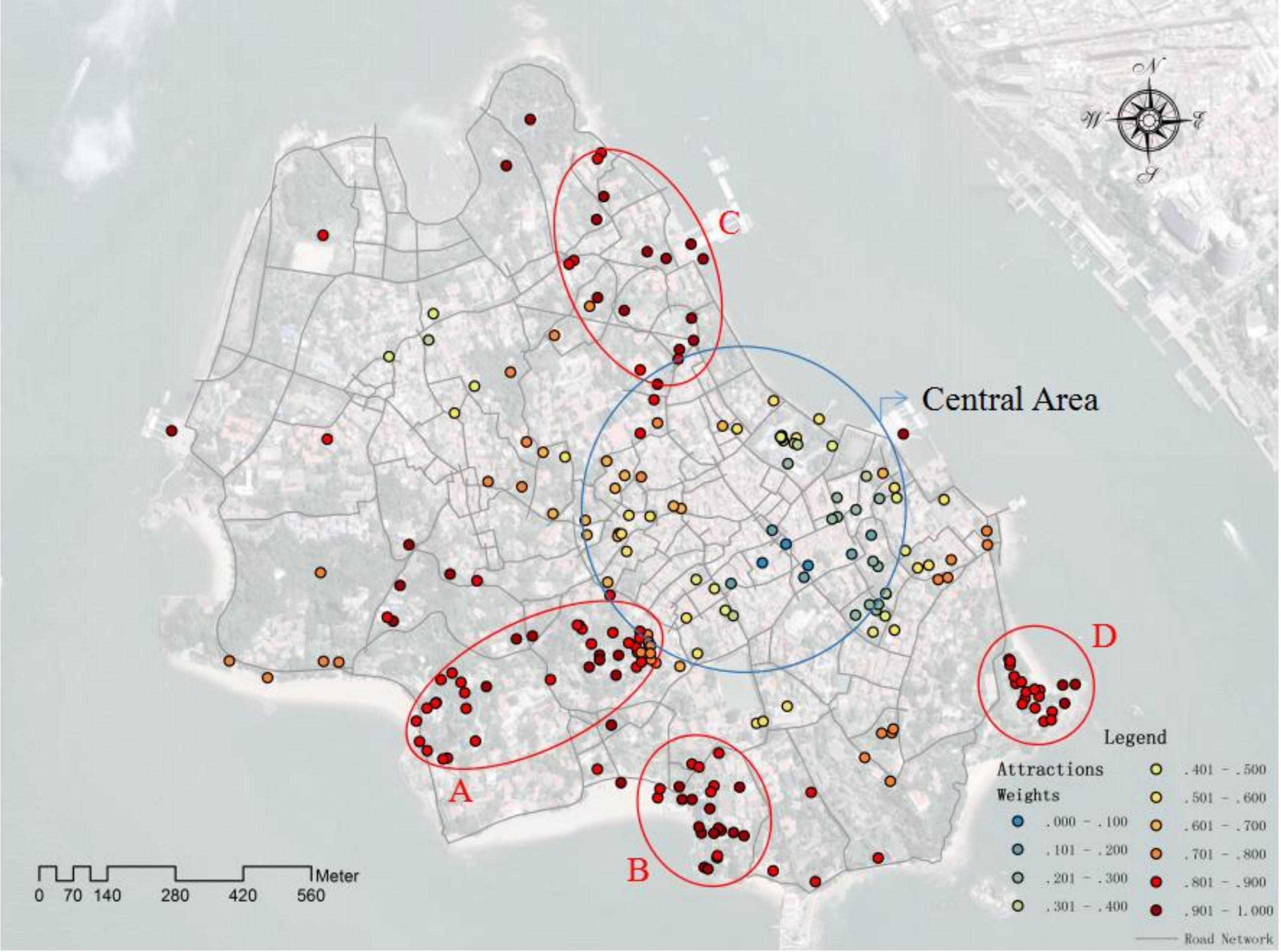

图 6　景点因子权重分布图

## 3.2 空间句法道路权重分析

### 3.2.1 距离类型与半径类型的选择

本研究以 $R^2$ 分析，检测距离选择度、拓朴选择度、角度选择度之间在选择不同半径大小的结果下的相关程度，结果如表 1 所示。

表 1　选择不同半径类型与半径大小之间的 $R^2$ 结果

| R= \ $R^2$ 类型 | 距离-拓朴 $R^2$ | 距离-角度 $R^2$ | 角度-拓朴 |
|---|---|---|---|
| 50 | $R^2$ =0.899276 | $R^2$ =0.75516 | $R^2$ =0.740829 |
| 100 | $R^2$ =0.868473 | $R^2$ =0.871189 | $R^2$ =0.807123 |
| 200 | $R^2$ =0.818688 | $R^2$ =0.862593 | $R^2$ =0.817587 |
| 300 | $R^2$ =0.79128 | $R^2$ =0.810974 | $R^2$ =0.814309 |
| 400 | $R^2$ =0.746457 | $R^2$ =0.753094 | $R^2$ =0.784453 |
| 500 | $R^2$ =0.688711 | $R^2$ =0.693626 | $R^2$ =0.720072 |
| 600 | $R^2$ =0.633049 | $R^2$ =0.648709 | $R^2$ =0.661987 |
| 800 | $R^2$ =0.547498 | $R^2$ =0.605844 | $R^2$ =0.596979 |
| 1000 | $R^2$=0.474125 | $R^2$=0.574587 | $R^2$=0.51216 |
| N | $R^2$=0.147846 | $R^2$=0.329615 | $R^2$=0.290651 |

数据显示三者中两两之间的相关程度，随着半径 R 的增加呈现负相关的情形，当 R 介于 50~400 米间，两两之间的 $R^2$ 皆呈现高度相关的情形，当 R 介于 500~1000 米间，两两之间的 $R^2$ 皆呈现中度相关的情形，而当 R 以全局尺度 n 作为考虑时，两两之间的 $R^2$ 则呈现低度相关的情形。虽然当 R=100 时，三者之间相较于其他半径会具有较高的相关程度，然而由过往对步行区的研究发现，步行 5 分钟 400 米半径的路程最常被做为评估步行区的依据(找研究)，因此本研究在 R 介于 50 至 400 米间选定以 400 米作为分析依据。图 7 为距离选择度、角度选择度、拓朴选择度在半径 400 米时的分析结果图。

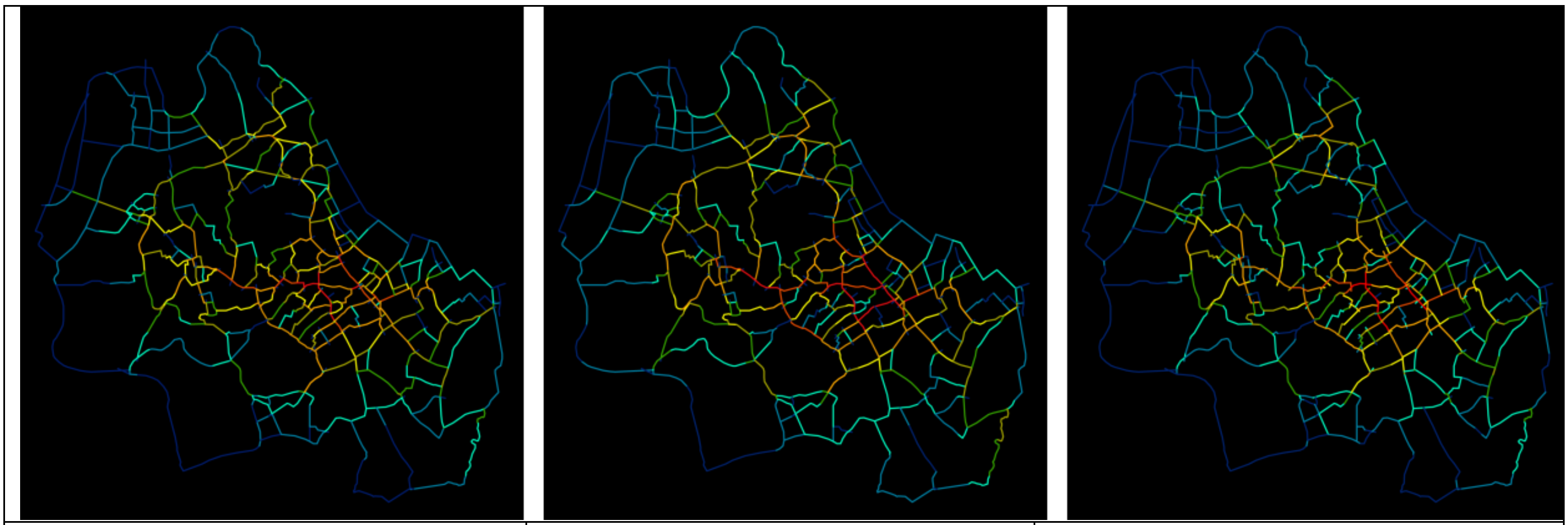

| 距离选择度 | 角度选择度 | 拓朴选择度 |
|---|---|---|

图 7 选择度分析图(R=400)

由表 1 及图 7 可知在 R=400 时，距离选择度、角度选择度、拓朴选择度的结果将呈现高度相关性，因此，本研究在后续的分析中将就其中之一进行分析。后续的章节中将会对其分析结果进行标准化以作为街道拥挤度参数使用。若进一步考虑三种半径模式的差异，距离模式由于评估两线段间的最短距离因此较具有争议，而本研究期望评估的是游客的步行模式，对不熟习当地环境的游客来说，单究道路类型而言，游客会偏向转折次数较少的方式进行游走以避免迷失方向，所以以计算最小转折角的角度模式则较不适用，而拓朴模式则能更好预测游客的步行结果。本研究将以拓朴选择度的分析结果进行后续分析。

### 3.2.2 道路空间句法的分析结果

根据 depthmap 软件进行拓朴选择度的街道拥挤度分析，结果如下图 8 所示，本研究共计分析 1,614 条线段，对应于总长 33,894 米的线段总长，本研究将所有数据依据选择度高低分为十个层级，如颜色结果越红则表示整合度越高，亦即作为串连交通景点的交通要道的能力越强；反之，若颜色越接近深蓝，则表示人群在选择路径时，较不容易借由该路径抵达其他区域。

以古城全区路网分布看选择度结果会发现越往古城中心区域选择度越高，且在道路越密集的区域选择度越高，显示越往中心这些道路越容易被行人穿越的情形。若仅以道路穿越的可能性可大致上分为双核心，主要核心为结廓 A，及次要核心结廓 B，主核心及次核心由一条横向道路串连在一起，使该横向道路具有全区中最高的选择度，而由结廓 A 及结廓 B 环绕的内部多为中至高选择度，游客相当容易在其中穿梭，而结廓 A 及结廓 B 环绕的外围则有曲折的环状道路分布并与鼓浪屿延着道屿边缘的环状道路进行横向连结，道路密度变得稀疏，越靠近岛屿外环道路的区域则选择度越低，较不容易成为游客选择穿越的街道。

鼓浪屿的道路拥挤度大致上可分为 5 个层级的结廓，依照选择度高到低可分为结廓 A、结廓 B、结廓 C、结廓 D 及结廓 E 等五种结廓类型，结廓 A 为全区主核心，位置接近岛屿中心易于穿越且街道密集交叉分布，导致了较高的选择度产生，其次为结廓 B，位置同样接近岛屿中心易于穿越，然而街道密度较低，因此整体选择度小于结廓 A 而成为次核心，结廓 C 及结廓 D 虽然位于核心区域外围，却由于邻近结廓 A 而有较高的整合度，而结廓 C 本身由于路往分布较密集因此整体选择度略高越于结廓 D，最后是位于外环区域的结廓 E，由于位于岛屿边缘地带，周边道路稀疏且因为位于边缘被选择

穿越的机会较少，导致了整体选择度最低的情形发生。

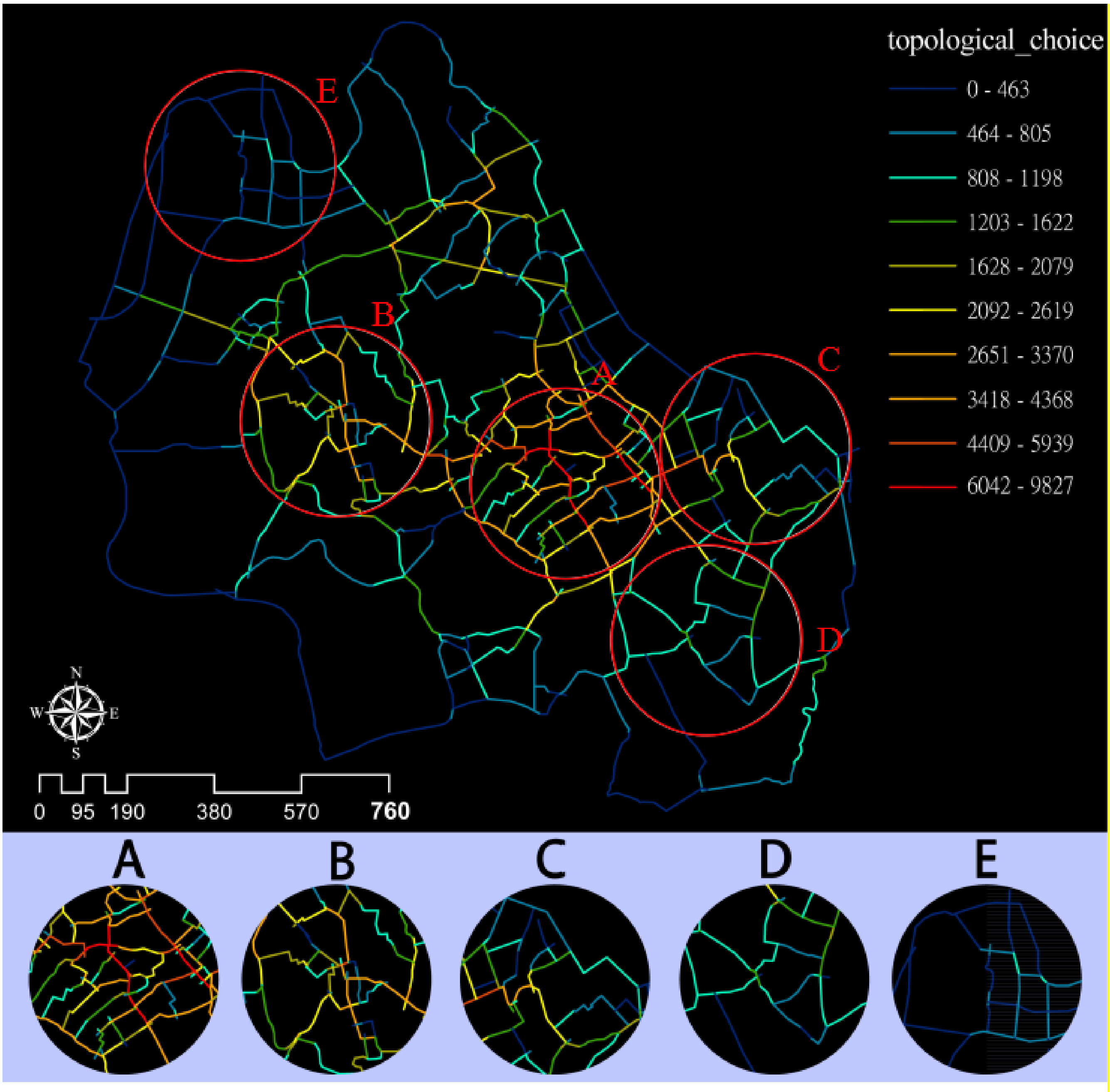

图 8 街道网络类型分布图

## 3.3 道路拥挤度模型的评估结果

根据上文所得到的道路可达性和景点吸引力权重，通过公式（1）和公式（2），我们可以得到道路的拥挤度情况。并在 ArcGIS 软件中进行处理与制图得到鼓浪屿岛最终的路网整合景点吸引力影响后的拥挤度图如图 9 所示。首先，道路整体的拥挤度，呈现中间密集，越往外拓展路网越稀疏的情况，我们知道，岛内居民主要生活于岛内中心部分，而旅游景点主要分布于岛的四周，这样的结果，可以看出，道路网仍以服务当地居民为主，服务旅游景点为辅。其次，浪屿岛中心部分 A 的路网多呈现为红色橘色黄色，说明中心部分道路拥挤度值比较高，鼓浪屿岛的中心部分道路较为拥挤；岛四周的路网颜色多为蓝色、青色、淡黄色，说明鼓浪屿环岛道路的路况并不拥挤，尤其是北侧部分 B 和西侧部分 C 道路拥挤度值较低。

我们结合景点的分布，对其进行进一步的分析，结果如图 10 所示。首先，我们发现，道路的拥挤度和景点的分布在空间上并没有很大的相关性。景点多的地方，道路大多呈现蓝色、青色、淡黄色的情况，说明鼓浪屿的大部分景点拥挤度不高，通行效果好。其次，中心区内的景点，道路的拥挤度较高，对景点的保护会有较大的压力。第三，仍有部分区域会因为景点的影响，导致道路拥挤

的情况，比如，在结廓 A 内，由于景点的大量分布，使得道路的拥挤度上升。最后，结廓 B 和结廓 C 内有大量较高权重的景点，但路网分布较为单一，只有一条可通行的道路，这可能使得道路受阻时，而无法到达的状况，因此，后期可以加大路面较宽或增加路网的条数，以缓解这种情况的发生。鼓浪屿道路的拥挤度分析有利于岛上路网的系统调整与规划，特别是缓解岛中心景点周边压力和历史景点的保护上。

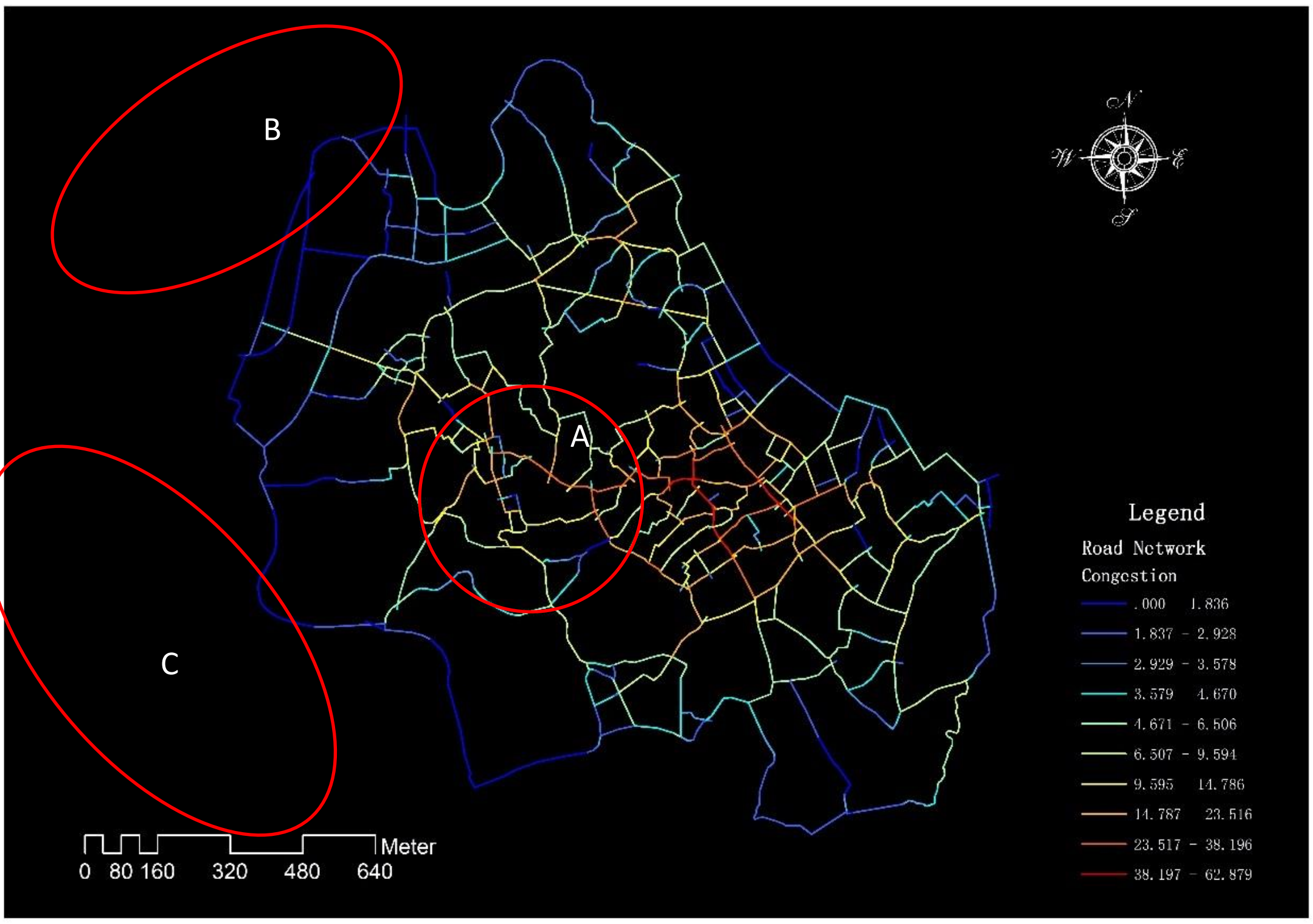

图 9 结合景点吸引力因子权重的道路拥挤度分析图 a

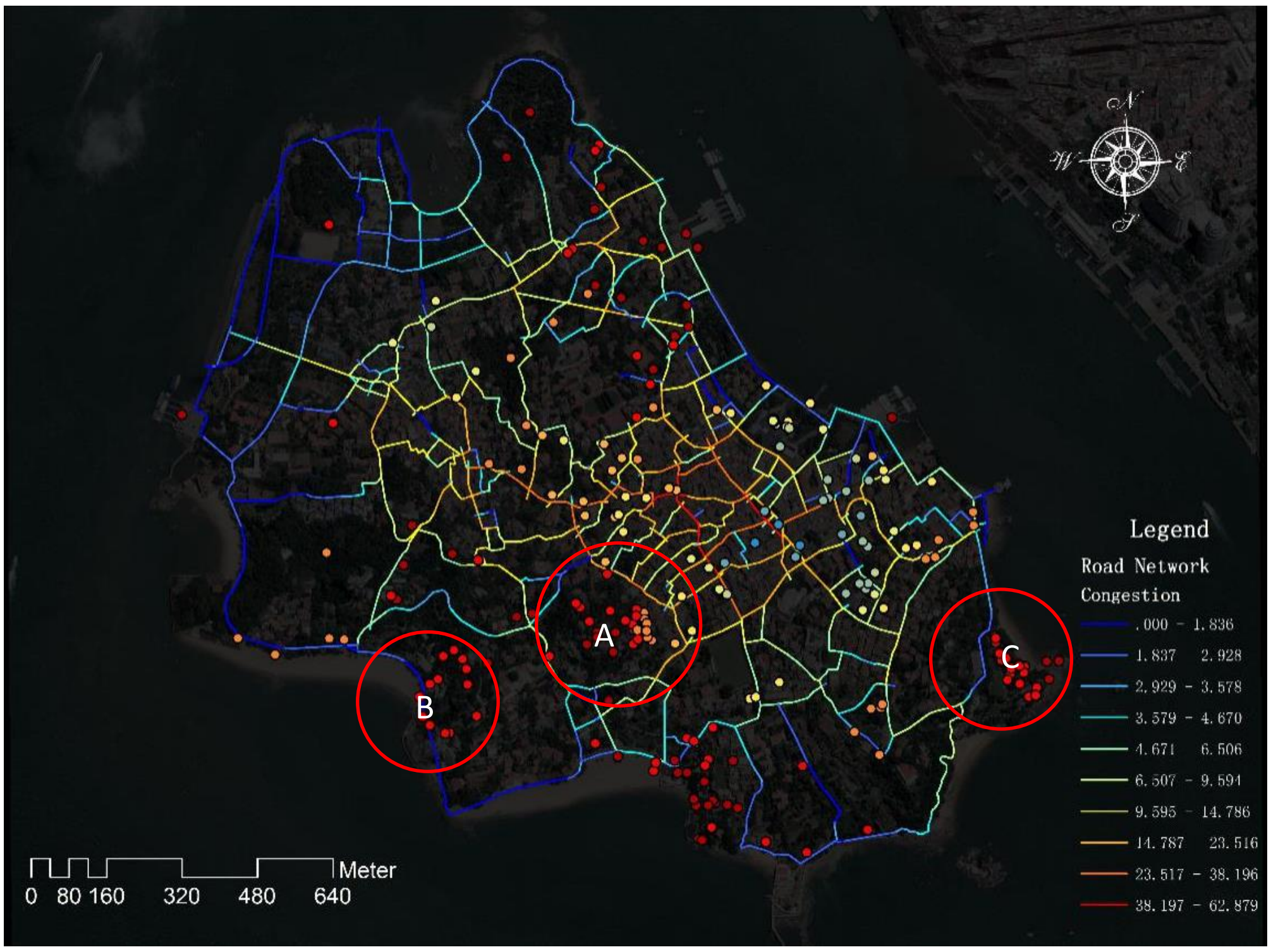

图 10 结合景点吸引力因子权重的道路拥挤度分析图 b

## 4. 讨论

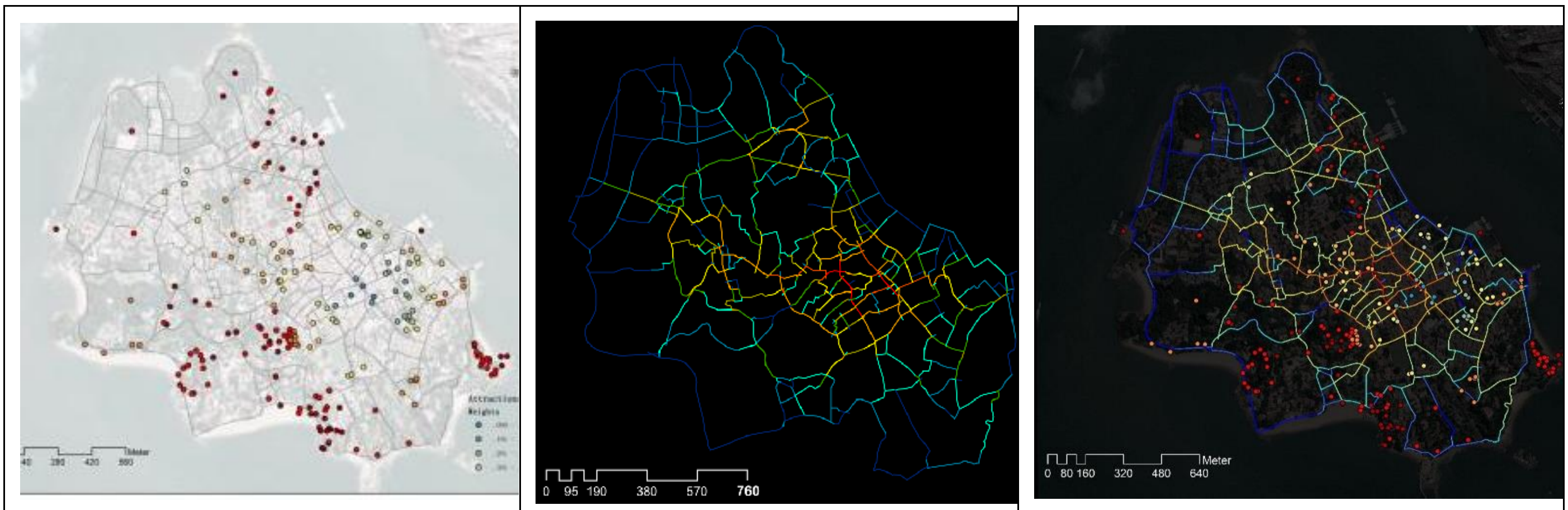

| 神经网络景点分布与权重 | 空间句法道路可达性 | 道路拥挤度 |
|---|---|---|

本研究试图发展空间句法分析的研究，并结合神经网络方法，以指导可持续旅游业的发展。一般而言，在道路的拥挤度评估模型中我们将街道网络结构，运动流量和景点分布之间进行关联，整体上可以预测较准确的行人流量。

我们的研究结果可以显着的发现他们三者的不同（图）。从图-A 我们只能在空间上知道有吸引力的旅游空间分布，但不能预测他们会多大程度引起道路的拥挤。图-B 是空间句法的分析结果，它能够很好的反映道路的可达性，但它不能很好的预测行人的偏好，因为游客很大程度上会受景点的吸引，而使得旅游的路径发生变化。我们建构的道路的拥挤度评估模型，可以整合街道网络结构和景点分布对行人运动流量的综合影响，可以在街道网络结构、景点分布和行人运动之间的对应性较低且不一致的情况下估算道路拥挤度的能力(Ye and van Nes, 2014)。值得注意的是，图-B 和图-C 两个结果之间存在一定的相似性。皮尔逊相关性进一步表明，道路拥挤度与空间句法的道路可达性之间存在较高的正相关系数（0.72）。换句话说，我们建构的模型很大程度上是居于空间句法的分析结果，但我们有对其进行了扩展，使得在评估阶段的可预测性相对更为成功。

我们从局部来看，我们发现图-B 和图-C 的中心整合度有不一致的现象。图-B 更加的往路网中心整合，代表着路网中心会有较高的拥挤情况；图-C 的中心整合度是相对分散的，说明行人的流量是相对分散，城市的中心性降低，也说明景点的吸引力可以有效的分散人群，进而反映出旅客的偏好。这样的结果，可以支持对旅游空间管理和历史建筑保护的研究。

## 5. 结论

在本文中，尝试了将空间结构、重要景点等因素，整合于同一个模型中来进行模拟。在空间句法和神经网络方法的框架下探讨城市街道网络结构和景观节点的吸引力对道路拥挤度的综合效应。客观、全面地研究这些携带着丰富历史信息的物质空间，可能造成拥挤度的后果，为历史街区得到更科学合理的发展奠定理论与实验基础。本文总体可归纳为以下几点：

首先，建构了对历史古城道路拥挤度的评估方法，可以提高预测旅客流量分布的能力以解决街道网络结构和景点分布对行人运动流量的综合影响，使历史古城不至于在实施或管理中造成大的错误。

其次，通过神经网络分析，我们知道，景点吸引力较高的地区主要分布在岛屿的边沿，这间接的反映了沿海的旅游资源较丰富；同时有卖门票的几个景点吸引力都达到 0.9 以上，也可以间接反映出神经网络分析的合理性；而城镇中心区的旅游景点的影响权重普遍不高，这也反映出中心区的旅游资源较弱。

再次，本文通过空间句法的三种距离类型评估，我们发现拓朴模式能够更好预测游客的步行结果。空间语法分析可以为我们提供历史城镇总体空间结构的准确说明。仅以道路穿越的可能性可大致上分为双核心，越往古城中心区域选择度越高，越靠近岛屿外环道路的区域则选择度越低。

最后，我们综合考虑了道路可达性和景点吸引力权重之间的相互影响。我们发现，道路的拥挤度和景点的分布在空间上并没有很大的相关性。中心区内的景点，道路的拥挤度较高，对景点的保护会有较大的压力。岛的四周的景点，拥挤度不高，但路网分布较为单一。

本研究仍有许多研究的不足。（Ⅰ）未考虑开阔空间和道路宽度对拥挤度的影响。开阔空间和道

路宽度会影响人口的承载力，因此后期研究可研拟开阔空间可容纳的人口数量和道路宽度对行人的影响。（Ⅱ）本研究在指标的考虑上主要是根据现状的影响因子进行考虑，而未考虑各指标之间的相互影响。在往后的研究中可以考虑指标之间的相互影响和景点之间的等级差别，而景点之间的等级差别可根据国家旅游景点等级的划分为主要依据。（Ⅲ）负面影响要素未进行考虑，比如交通管制等。负面影响会影响景点的吸引力，会对模拟结果产生影响，因此后期研究有必要对其进行考虑。

## 参考文献

## 附表 鼓浪屿景点各影响因子的数量和景点吸引力

| 景点名称 | 200 米范围内影响因子个数 | | | | | | | | 景点吸引力权重 |
|---|---|---|---|---|---|---|---|---|---|
| | 宾馆 | 餐饮 | 购物 | 周围景点 | 休闲娱乐 | 交通设施 | 生活服务 | 公共服务 | |
| Gulangyu | 28 | 14 | 5 | 19 | 2 | 12 | 5 | 10 | 0.648 |
| Xiamen Gulangyu Wharf | 10 | 11 | 13 | 10 | 1 | 6 | 12 | 2 | 0.978 |
| Sunlight rock | 16 | 7 | 6 | 33 | 5 | 11 | 9 | 1 | 0.960 |
| The most beautiful corner | 25 | 4 | 2 | 11 | 2 | 8 | 8 | 0 | 0.977 |
| Xinzhuang Garden | 5 | 3 | 1 | 24 | 4 | 5 | 11 | 0 | 0.917 |
| Sanqiutian Wharf | 13 | 4 | 2 | 6 | 1 | 8 | 8 | 0 | 0.981 |
| Gulangyu Inner Ring | 0 | 4 | 2 | 0 | 0 | 2 | 6 | 1 | 0.923 |
| Gulangyu Organ Museum | 33 | 10 | 13 | 14 | 2 | 6 | 11 | 3 | 0.978 |
| Haoyueyuan | 11 | 3 | 0 | 18 | 3 | 2 | 1 | 0 | 0.870 |
| Xiamen Underwater World | 33 | 64 | 46 | 20 | 4 | 11 | 24 | 13 | 0.533 |
| Zheng Chenggong Memorial Hall | 17 | 4 | 1 | 33 | 3 | 11 | 10 | 5 | 0.837 |
| Zheng Chenggong Stone Statue | 0 | 0 | 0 | 18 | 1 | 1 | 1 | 0 | 0.910 |
| Shell dream world | 3 | 3 | 4 | 15 | 4 | 3 | 7 | 0 | 0.856 |
| Gulangyu-Beach | 0 | 2 | 4 | 14 | 4 | 2 | 7 | 0 | 0.842 |
| Sea paradise | 34 | 22 | 21 | 21 | 8 | 6 | 10 | 16 | 0.231 |
| Gulangyu rare world | 49 | 49 | 32 | 18 | 4 | 10 | 20 | 12 | 0.559 |
| Street Heart Park (Fenglong Road) | 57 | 91 | 87 | 24 | 11 | 22 | 27 | 26 | 0.172 |
| Centennial Gulangyu Museum | 42 | 70 | 80 | 25 | 12 | 16 | 21 | 23 | 0.096 |
| U.S. Consulate | 18 | 4 | 2 | 12 | 1 | 8 | 8 | 0 | 1.000 |
| Pen Mountain Cave | 50 | 10 | 6 | 7 | 1 | 4 | 2 | 5 | 0.769 |
| Qinyuan | 6 | 6 | 4 | 17 | 3 | 5 | 10 | 0 | 0.924 |
| World Celebrity Wax Figure 3D Art Gallery | 10 | 2 | 3 | 9 | 0 | 4 | 4 | 5 | 0.774 |
| Xiamen Underwater World - Aquarium | 36 | 65 | 49 | 18 | 3 | 11 | 25 | 15 | 0.498 |
| Forbidden City Gulangyu Foreign Cultural Relics Museum | 8 | 1 | 1 | 5 | 1 | 1 | 2 | 0 | 0.890 |
| Gulangyu Fun Wax Museum | 16 | 36 | 40 | 27 | 2 | 9 | 24 | 20 | 0.364 |
| Xiamen Gulangyu Scenic Area - Catholic Church | 34 | 40 | 41 | 32 | 7 | 7 | 21 | 17 | 0.295 |
| Gossip | 24 | 6 | 10 | 12 | 2 | 6 | 11 | 2 | 0.891 |
| Gulangshi | 7 | 6 | 1 | 3 | 4 | 2 | 2 | 1 | 0.764 |
| Trinity Hall | 29 | 12 | 5 | 21 | 2 | 13 | 5 | 9 | 0.697 |
| Gion | 13 | 8 | 7 | 4 | 2 | 4 | 5 | 3 | 0.782 |
| HSBC Bank House Site | 31 | 8 | 1 | 9 | 3 | 9 | 4 | 4 | 0.789 |
| Gulangyu International Lettering Art Museum | 5 | 6 | 5 | 26 | 5 | 6 | 12 | 1 | 0.863 |
| Hong Kong Aberdeen Beach | 4 | 4 | 4 | 26 | 5 | 3 | 10 | 0 | 0.866 |
| The former site of the Japanese consulate | 24 | 46 | 45 | 31 | 4 | 7 | 21 | 25 | 0.177 |
| Longshan Cave | 23 | 5 | 10 | 12 | 0 | 7 | 13 | 2 | 0.961 |
| Kindergarten | 30 | 15 | 10 | 4 | 4 | 4 | 4 | 11 | 0.440 |
| Pen Mountain Park | 40 | 11 | 5 | 9 | 0 | 8 | 2 | 8 | 0.718 |
| Spring Cottage | 40 | 10 | 5 | 9 | 1 | 6 | 0 | 5 | 0.774 |
| Xiamen Ocean Park | 25 | 43 | 39 | 22 | 1 | 10 | 26 | 18 | 0.456 |
| Fanpo Building | 37 | 17 | 12 | 21 | 4 | 16 | 9 | 11 | 0.616 |
| Fall and fall | 60 | 47 | 53 | 26 | 16 | 21 | 14 | 9 | 0.421 |
| Gulangyu Concord Chapel | 36 | 54 | 59 | 31 | 6 | 10 | 22 | 24 | 0.180 |
| Xiamen Oriental Fishbone Art Museum | 44 | 25 | 19 | 17 | 3 | 12 | 12 | 6 | 0.791 |
| Former British Consulate | 19 | 18 | 27 | 20 | 2 | 8 | 21 | 13 | 0.545 |
| Yanweishan Ecological Park | 3 | 0 | 0 | 3 | 0 | 1 | 1 | 0 | 0.906 |
| Gulangyu·Shell Museum | 3 | 3 | 4 | 15 | 4 | 3 | 7 | 0 | 0.856 |
| Guanhaiyuan Beach | 0 | 0 | 0 | 3 | 0 | 1 | 0 | 2 | 0.823 |
| Gong Baodi | 30 | 8 | 1 | 9 | 3 | 7 | 2 | 3 | 0.800 |
| Junyi Beach | 0 | 2 | 4 | 14 | 4 | 0 | 7 | 0 | 0.824 |
| Xiamen Overseas Chinese Subtropical Plant Introduction Park - Ornamental Plant Area | 4 | 3 | 1 | 14 | 2 | 4 | 5 | 0 | 0.917 |
| Guan Cai Building | 35 | 12 | 6 | 13 | 0 | 5 | 3 | 8 | 0.700 |
| Drum sound hole | 0 | 2 | 4 | 14 | 4 | 0 | 8 | 0 | 0.828 |
| Gulangyu Guanhai Park | 5 | 4 | 0 | 12 | 3 | 5 | 6 | 2 | 0.820 |
| Revitalizing Gulangyu Sanhe Gongji Cliff Stone Carvings | 30 | 7 | 1 | 11 | 3 | 10 | 5 | 0 | 0.958 |
| Lin Qiaozhi Memorial Hall | 16 | 9 | 7 | 4 | 3 | 4 | 5 | 3 | 0.755 |
| Former German consulate ruins | 20 | 17 | 28 | 25 | 2 | 9 | 20 | 15 | 0.498 |
| Huang Simin Villa | 41 | 19 | 9 | 30 | 7 | 18 | 10 | 11 | 0.579 |
| Beacon Hill | 33 | 12 | 4 | 7 | 4 | 5 | 2 | 4 | 0.710 |
| Yangjiayuan | 40 | 15 | 15 | 15 | 3 | 10 | 10 | 3 | 0.873 |
| An Xiantang | 22 | 3 | 5 | 6 | 0 | 4 | 2 | 0 | 0.958 |
| China Record Museum | 30 | 15 | 21 | 26 | 6 | 7 | 12 | 16 | 0.316 |
| Sunlight Rock Temple | 23 | 9 | 5 | 35 | 5 | 9 | 7 | 4 | 0.794 |
| Former court trial | 33 | 12 | 3 | 18 | 2 | 13 | 4 | 9 | 0.692 |
| Ma John Memorial | 7 | 9 | 6 | 11 | 7 | 7 | 7 | 6 | 0.566 |
| Gulangyu Pien Tze Huang Museum | 44 | 20 | 19 | 17 | 3 | 11 | 11 | 5 | 0.816 |
| Xiamen Overseas Chinese Subtropical Plant Introduction Park | 5 | 4 | 0 | 10 | 3 | 3 | 4 | 0 | 0.866 |
| Gulangyu Food Factory | 53 | 29 | 23 | 20 | 5 | 17 | 13 | 8 | 0.723 |
| Sunlight Rock - Peak | 13 | 7 | 5 | 32 | 3 | 11 | 9 | 1 | 0.975 |
| Former Residence of Lin Zumi | 29 | 8 | 1 | 8 | 3 | 8 | 2 | 3 | 0.806 |
| Gulangyu Engineering Bureau Site | 26 | 7 | 10 | 13 | 1 | 7 | 15 | 2 | 0.947 |
| Twelve holes | 5 | 4 | 3 | 26 | 4 | 5 | 13 | 1 | 0.890 |
| Demeanor | 67 | 55 | 37 | 19 | 10 | 17 | 13 | 6 | 0.649 |
| Yanweishan Noon Site | 1 | 0 | 1 | 3 | 0 | 2 | 0 | 0 | 0.908 |
| Huang Rong Yuan Tang Villa | 30 | 15 | 21 | 26 | 6 | 7 | 12 | 16 | 0.316 |
| Lin Qiaozhi doctor statue | 14 | 8 | 7 | 5 | 2 | 4 | 5 | 3 | 0.786 |
| Xu Jiayuan | 37 | 18 | 8 | 5 | 4 | 5 | 4 | 14 | 0.361 |
| Ice and snow | 25 | 54 | 54 | 30 | 6 | 12 | 28 | 23 | 0.231 |
| Zhujia Garden | 43 | 38 | 41 | 22 | 15 | 18 | 9 | 10 | 0.349 |
| Ningyuan Building | 36 | 18 | 9 | 18 | 6 | 17 | 7 | 9 | 0.625 |
| Gulangyu British Consular Residence | 7 | 3 | 0 | 6 | 3 | 1 | 2 | 2 | 0.755 |
| Gospel hall | 18 | 10 | 12 | 32 | 9 | 9 | 7 | 5 | 0.619 |
| Doctor | 60 | 39 | 44 | 34 | 14 | 20 | 11 | 11 | 0.411 |
| Xiamen Gulangyu Tourist Area Management Office | 16 | 6 | 6 | 35 | 4 | 5 | 7 | 5 | 0.742 |
| Gulangyu Historical and Cultural Exhibition Hall | 15 | 23 | 30 | 22 | 2 | 7 | 22 | 11 | 0.607 |
| Xiamen Underwater World - Undersea Tunnel | 33 | 62 | 42 | 16 | 3 | 9 | 23 | 15 | 0.470 |
| Fuxing Hall | 24 | 9 | 11 | 23 | 3 | 5 | 7 | 4 | 0.785 |
| Gulangdongtian | 17 | 6 | 6 | 32 | 3 | 5 | 8 | 2 | 0.884 |
| Drum Tower | 18 | 7 | 6 | 35 | 4 | 8 | 7 | 5 | 0.771 |
| Ruins of Gulangyu Island Scenic Spot in Gulangyu Island, Xiamen | 11 | 4 | 2 | 7 | 1 | 8 | 8 | 0 | 0.981 |
| Yizu Villa | 27 | 11 | 4 | 16 | 1 | 8 | 5 | 11 | 0.599 |
| The former site of the CPC Fujian Provincial Committee | 43 | 18 | 22 | 32 | 11 | 13 | 7 | 7 | 0.542 |
| Sperm whale specimen museum | 46 | 73 | 68 | 21 | 5 | 14 | 25 | 24 | 0.255 |
| Country name well | 6 | 5 | 5 | 29 | 2 | 3 | 11 | 1 | 0.929 |
| Meize Building | 68 | 47 | 33 | 19 | 9 | 15 | 12 | 6 | 0.660 |
| Yinghua Gate Building | 42 | 22 | 14 | 20 | 5 | 17 | 11 | 11 | 0.606 |
| Zhaohe Mountain Park | 2 | 0 | 0 | 0 | 0 | 1 | 3 | 2 | 0.830 |
| Lu Yuzhang | 8 | 7 | 1 | 5 | 5 | 2 | 3 | 1 | 0.743 |
| Hutchison Ocean ruins | 22 | 4 | 6 | 12 | 0 | 8 | 10 | 1 | 0.998 |
| Ancient summer shelter | 17 | 6 | 6 | 33 | 4 | 9 | 9 | 1 | 0.935 |
| London Missionary Missionary House | 7 | 5 | 1 | 18 | 3 | 7 | 7 | 0 | 0.934 |
| Longevity Park | 52 | 20 | 13 | 7 | 4 | 5 | 3 | 10 | 0.506 |
| Tomb of Chen Shijing | 17 | 6 | 0 | 8 | 2 | 9 | 4 | 2 | 0.879 |
| Xiamen story | 28 | 12 | 16 | 28 | 9 | 10 | 5 | 6 | 0.580 |
| Yin Chengzong | 17 | 4 | 1 | 6 | 1 | 4 | 2 | 0 | 0.927 |
| Forty four bridge | 4 | 3 | 1 | 25 | 4 | 5 | 11 | 0 | 0.918 |
| The former site of the Asian Asia Fire Oil Company | 44 | 39 | 42 | 24 | 15 | 18 | 10 | 8 | 0.423 |
| Edition Zhu Chuanfang | 36 | 18 | 9 | 20 | 6 | 17 | 8 | 9 | 0.633 |
| Former residence of Xu Feiping | 50 | 22 | 13 | 9 | 4 | 5 | 3 | 13 | 0.407 |
| Xingxian Palace | 23 | 9 | 11 | 21 | 3 | 5 | 6 | 5 | 0.738 |
| Hero rock | 1 | 2 | 4 | 14 | 3 | 0 | 8 | 0 | 0.859 |
| Xinzhuang Tourism Station | 5 | 6 | 5 | 24 | 5 | 5 | 12 | 1 | 0.849 |
| Yanping Park | 4 | 3 | 4 | 8 | 2 | 2 | 6 | 0 | 0.887 |
| Turtle | 36 | 64 | 45 | 18 | 3 | 9 | 25 | 15 | 0.481 |
| HSBC staff apartment apartment | 28 | 6 | 1 | 8 | 2 | 8 | 2 | 4 | 0.797 |
| One of the sea paradise | 31 | 13 | 13 | 21 | 8 | 5 | 9 | 9 | 0.455 |
| Huang Yuxiang's former residence | 61 | 36 | 23 | 20 | 8 | 21 | 10 | 12 | 0.535 |
| Huang's small | 61 | 58 | 57 | 13 | 16 | 20 | 11 | 15 | 0.194 |
| Three let go | 25 | 9 | 4 | 38 | 4 | 16 | 12 | 7 | 0.803 |
| Hero Park | 1 | 2 | 4 | 14 | 3 | 0 | 8 | 0 | 0.859 |
| Yanping Film Museum | 56 | 76 | 78 | 13 | 14 | 18 | 20 | 26 | 0.000 |
| Guanhai Promenade (Chuzhou Road) | 1 | 0 | 0 | 18 | 2 | 1 | 1 | 0 | 0.881 |
| Haoyuelou | 5 | 4 | 4 | 26 | 5 | 6 | 12 | 0 | 0.903 |
| Parrot show | 5 | 3 | 3 | 18 | 4 | 4 | 9 | 0 | 0.883 |
| Xilin·Qingqing Villa | 16 | 4 | 1 | 32 | 4 | 12 | 10 | 3 | 0.890 |
| Sea paradise | 33 | 14 | 17 | 20 | 8 | 7 | 9 | 9 | 0.472 |
| Salvation Hospital and Nurse School Site | 10 | 1 | 1 | 5 | 1 | 1 | 2 | 0 | 0.892 |
| Eucalyptus | 30 | 7 | 12 | 11 | 2 | 8 | 13 | 3 | 0.880 |
| Huangjiadu Wharf | 21 | 36 | 32 | 17 | 1 | 9 | 20 | 13 | 0.572 |
| Emperor temple | 0 | 0 | 0 | 18 | 1 | 0 | 1 | 0 | 0.901 |
| Pangu Museum | 36 | 43 | 44 | 32 | 7 | 7 | 21 | 18 | 0.267 |
| Bumper car playground | 4 | 5 | 4 | 26 | 5 | 6 | 10 | 0 | 0.893 |
| Octagonal building | 27 | 12 | 12 | 26 | 4 | 5 | 6 | 10 | 0.539 |
| True rate pavilion | 5 | 4 | 1 | 25 | 4 | 5 | 13 | 0 | 0.927 |
| Wandering Mountain | 10 | 5 | 0 | 6 | 4 | 3 | 3 | 1 | 0.788 |
| Japanese Police Department and Dormitory Site | 24 | 46 | 45 | 31 | 4 | 7 | 21 | 25 | 0.177 |
| Repin Art Gallery | 50 | 82 | 84 | 19 | 12 | 18 | 26 | 26 | 0.076 |
| Former site of the Southern Church | 41 | 68 | 73 | 25 | 13 | 16 | 17 | 21 | 0.101 |
| Piano gallery | 5 | 3 | 1 | 26 | 3 | 5 | 11 | 0 | 0.951 |
| Hongyi Master Memorial Hall | 19 | 6 | 5 | 33 | 4 | 6 | 8 | 3 | 0.830 |
| Gulangyu Piano Museum - Hall II | 5 | 3 | 1 | 26 | 3 | 5 | 11 | 0 | 0.951 |
| Zhongnan Bank's former site | 32 | 65 | 61 | 31 | 6 | 14 | 31 | 24 | 0.246 |
| Lin Wenqing Villa | 35 | 15 | 11 | 22 | 3 | 14 | 10 | 10 | 0.665 |
| Gulangyu Konka Amusement Park | 2 | 3 | 3 | 14 | 2 | 3 | 7 | 0 | 0.913 |
| The former site of the Xiamen Customs Deputy Taxation Department | 11 | 3 | 0 | 6 | 2 | 2 | 3 | 3 | 0.764 |
| Former residence of Lu Jiaxi | 37 | 19 | 9 | 18 | 6 | 17 | 7 | 11 | 0.557 |
| Climbing house | 4 | 3 | 3 | 19 | 4 | 6 | 8 | 0 | 0.898 |
| Wu Ancestral Hall | 14 | 10 | 8 | 0 | 2 | 4 | 4 | 1 | 0.845 |
| Xiamen Gulangyu Scenic Area - Nikko Temple | 23 | 9 | 3 | 35 | 5 | 9 | 8 | 3 | 0.838 |
| Xiamen Customs Communication Tower | 24 | 2 | 0 | 7 | 1 | 6 | 2 | 0 | 0.954 |
| Duyue Pavilion | 4 | 4 | 3 | 25 | 4 | 5 | 12 | 0 | 0.921 |
| Gulang Cave Photo | 1 | 2 | 4 | 15 | 4 | 2 | 7 | 0 | 0.845 |
| Former Danish Dabei Audio-visual Bureau | 1 | 0 | 0 | 5 | 0 | 1 | 2 | 1 | 0.875 |
| Yiyuan | 31 | 11 | 11 | 17 | 6 | 4 | 7 | 7 | 0.564 |
| Zheng Chenggong's screen | 1 | 0 | 0 | 18 | 2 | 1 | 1 | 0 | 0.881 |
| Light complex | 16 | 7 | 6 | 33 | 5 | 10 | 9 | 1 | 0.913 |
| Dongsheng arch | 14 | 2 | 3 | 8 | 0 | 4 | 4 | 5 | 0.775 |
| Beauty park | 32 | 12 | 22 | 31 | 4 | 6 | 15 | 14 | 0.453 |
| Yuantong Gate | 16 | 6 | 6 | 35 | 4 | 5 | 7 | 5 | 0.742 |
| Waterfowl | 5 | 4 | 5 | 16 | 5 | 4 | 8 | 0 | 0.843 |
| The address of the Great Northern Telegraph Office | 2 | 2 | 1 | 16 | 2 | 1 | 3 | 0 | 0.885 |
| Commercial office Xiamen Telephone | 34 | 67 | 65 | 31 | 6 | 13 | 31 | 24 | 0.236 |
| Sea paradise | 32 | 20 | 21 | 23 | 7 | 7 | 10 | 16 | 0.275 |
| Clock tower | 22 | 8 | 5 | 36 | 5 | 9 | 7 | 5 | 0.758 |
| The Twilight Drive bronze relief | 6 | 0 | 0 | 18 | 2 | 1 | 1 | 0 | 0.886 |
| Hongyi Master Memorial Park | 17 | 4 | 6 | 32 | 3 | 5 | 7 | 3 | 0.842 |
| Winning with the sun | 14 | 6 | 6 | 34 | 4 | 8 | 10 | 1 | 0.930 |
| Over the ancient well | 8 | 1 | 0 | 18 | 3 | 1 | 1 | 0 | 0.858 |
| Yuantong Hall | 20 | 6 | 4 | 35 | 4 | 5 | 8 | 3 | 0.828 |
| Tibetan temple | 22 | 8 | 5 | 36 | 5 | 9 | 7 | 5 | 0.758 |
| Qiuqiu | 5 | 4 | 2 | 24 | 4 | 5 | 12 | 0 | 0.920 |
| Clock building | 45 | 21 | 12 | 20 | 6 | 17 | 9 | 12 | 0.540 |
| Longtoushan ruins | 16 | 4 | 1 | 32 | 3 | 11 | 9 | 5 | 0.830 |
| Peacock table | 5 | 3 | 4 | 17 | 4 | 4 | 8 | 0 | 0.876 |
| Haoyuexiongfeng | 11 | 3 | 0 | 18 | 3 | 2 | 1 | 0 | 0.870 |
| Dream Medusa Exhibition | 35 | 62 | 45 | 16 | 3 | 9 | 24 | 15 | 0.473 |
| The second session of the Communist Party of China in Fujian Province | 23 | 13 | 7 | 4 | 1 | 4 | 2 | 13 | 0.455 |
| Mid-Autumn Festival Boa Statue | 17 | 7 | 5 | 32 | 5 | 12 | 9 | 2 | 0.892 |
| Pillow flow stone | 4 | 3 | 1 | 25 | 4 | 5 | 11 | 0 | 0.918 |
| Qianbo Pavilion | 2 | 2 | 1 | 21 | 2 | 1 | 7 | 0 | 0.914 |
| Chaoyin Pavilion | 0 | 0 | 0 | 18 | 2 | 1 | 1 | 0 | 0.880 |
| Bohai Xiongfeng | 2 | 0 | 0 | 18 | 2 | 1 | 1 | 0 | 0.882 |
| Water station ruins | 19 | 7 | 4 | 36 | 5 | 10 | 9 | 2 | 0.887 |
| Gulangyu Chinese Sculpture Art Harbor | 28 | 10 | 5 | 40 | 5 | 13 | 11 | 9 | 0.677 |
| Christian Chicken Mountain Road Party Point | 9 | 2 | 0 | 6 | 0 | 6 | 3 | 0 | 0.970 |
| Recruiting pavilion | 2 | 2 | 1 | 21 | 2 | 1 | 7 | 0 | 0.914 |
| Yi Yun Building | 6 | 3 | 5 | 15 | 3 | 0 | 8 | 0 | 0.866 |
| Xiamen Customs Ship Management Office | 24 | 3 | 0 | 10 | 3 | 6 | 3 | 0 | 0.907 |
| Zheng Chenggong Micro-carving Exhibition Hall | 2 | 0 | 0 | 18 | 2 | 1 | 1 | 0 | 0.882 |
| Linerjia bronze statue | 5 | 4 | 3 | 26 | 4 | 5 | 13 | 1 | 0.890 |
| Former Japanese Pok Oi Hospital | 19 | 6 | 10 | 19 | 2 | 9 | 11 | 11 | 0.596 |
| Walking in the poultry garden | 4 | 3 | 3 | 17 | 4 | 3 | 7 | 0 | 0.862 |
| Gulangyu Piano Museum - Hall 1 | 4 | 3 | 1 | 23 | 3 | 1 | 7 | 0 | 0.891 |
| Serpentine theme exhibition area | 37 | 67 | 51 | 20 | 4 | 11 | 24 | 15 | 0.473 |
| Shenjia Tea Garden | 50 | 43 | 32 | 22 | 4 | 13 | 19 | 10 | 0.661 |
| Xiamen Gulangyu Scenic Area - Forest House (suspended) | 8 | 5 | 1 | 20 | 3 | 10 | 7 | 1 | 0.928 |
| Wuzhou Conference Hall | 5 | 4 | 2 | 28 | 3 | 5 | 13 | 1 | 0.925 |
| Guanhai Quqiao | 0 | 0 | 0 | 18 | 1 | 1 | 1 | 0 | 0.910 |
| Slan International Cultural Exchange Center | 31 | 27 | 23 | 13 | 2 | 9 | 14 | 11 | 0.597 |
| Sunlight Rock Temple Service Department | 22 | 7 | 5 | 35 | 4 | 9 | 7 | 5 | 0.785 |
| Listen to Tao Xuan | 4 | 3 | 1 | 23 | 3 | 1 | 7 | 0 | 0.891 |
| Haizang Pavilion | 1 | 0 | 0 | 18 | 2 | 1 | 1 | 0 | 0.881 |
| Feng Taoting | 0 | 0 | 0 | 18 | 2 | 1 | 1 | 0 | 0.880 |
| MIAD Factory (Sanyou Holiday Tourism City) | 30 | 65 | 58 | 30 | 6 | 15 | 29 | 25 | 0.230 |
| Sea paradise | 36 | 27 | 27 | 24 | 7 | 7 | 13 | 16 | 0.289 |
| Hard Rock Mountain House | 6 | 6 | 5 | 22 | 5 | 5 | 12 | 1 | 0.846 |
| Heart pavilion | 0 | 0 | 0 | 18 | 2 | 1 | 1 | 0 | 0.880 |
| South monument | 1 | 0 | 0 | 18 | 2 | 1 | 1 | 0 | 0.881 |
| Mr. Hu Youyi | 4 | 3 | 1 | 23 | 3 | 1 | 9 | 0 | 0.899 |
| Special Aquatic Exhibition Area | 35 | 63 | 44 | 17 | 3 | 9 | 23 | 15 | 0.473 |
| Xiamen Gulangyu Xingxian Palace Council | 22 | 4 | 4 | 22 | 3 | 4 | 6 | 5 | 0.735 |
| Terraced hills | 7 | 6 | 3 | 24 | 5 | 6 | 9 | 0 | 0.887 |
| Huang Jude Hall | 29 | 6 | 3 | 17 | 3 | 10 | 9 | 1 | 0.949 |
| Xiting | 5 | 3 | 1 | 24 | 4 | 5 | 11 | 0 | 0.917 |
| Nine summers | 20 | 6 | 4 | 35 | 4 | 5 | 9 | 2 | 0.870 |
| Daxiong Hall | 18 | 7 | 5 | 34 | 5 | 6 | 7 | 3 | 0.798 |
| Zhangzhou portrait | 11 | 3 | 0 | 18 | 3 | 2 | 1 | 0 | 0.870 |
| Bushan Pavilion | 4 | 3 | 1 | 24 | 4 | 5 | 11 | 0 | 0.916 |
| Lin's House - Conference Hall | 28 | 12 | 17 | 27 | 4 | 6 | 8 | 10 | 0.556 |
| Omiya Post Inspection Building | 14 | 8 | 10 | 6 | 7 | 8 | 4 | 5 | 0.593 |
| blend | 20 | 30 | 35 | 22 | 2 | 8 | 24 | 23 | 0.274 |
| Amitabha Temple | 21 | 8 | 4 | 36 | 5 | 6 | 8 | 3 | 0.810 |
| Sea paradise | 39 | 30 | 34 | 19 | 8 | 8 | 11 | 16 | 0.252 |
| Ocean Science Knowledge Gallery | 35 | 66 | 54 | 20 | 3 | 11 | 24 | 20 | 0.359 |
| Gulangyu Island - Tourists in Xiaotun District | 7 | 6 | 1 | 2 | 4 | 1 | 2 | 1 | 0.752 |
| The Great Navigation Age and Gulangyu-Western Ancient Literature Exhibition | 6 | 9 | 8 | 10 | 8 | 7 | 7 | 5 | 0.569 |
| Galan Palace | 18 | 7 | 6 | 35 | 4 | 8 | 7 | 5 | 0.771 |
| Chen Jinfang's former residence | 8 | 6 | 1 | 6 | 5 | 2 | 3 | 1 | 0.746 |
| Yong Cui Ting | 1 | 0 | 0 | 18 | 2 | 1 | 1 | 0 | 0.881 |
| Yue Boting | 1 | 0 | 0 | 18 | 2 | 1 | 1 | 0 | 0.881 |